\documentclass[runningheads]{llncs}
\usepackage{graphicx}
\usepackage{amsmath}
\usepackage{bookmark}
\usepackage{times}
\usepackage{xcolor}
\usepackage{verbatim}
\usepackage[utf8]{inputenc}
\usepackage{hyperref}
\usepackage{amssymb}

\usepackage{amsthm}
\usepackage{listings}
\usepackage{nameref}
\usepackage{svg}
\usepackage{calc}
\usepackage{newfloat}
\usepackage{enumitem}
\usepackage{subfig}
\usepackage{float}
\usepackage{makecell}
\usepackage[newfloat,frozencache,cachedir=_minted-main]{minted}
\usepackage{csquotes}

\begin{document}
\title{SIGNAL - The SAP Signavio Analytics Query Language}
%
\author{Timotheus Kampik \and Andre Lücke \and Jörn Horstmann \and Mark Wheeler \and David Eickhoff}

\institute{
    SAP Signavio, Berlin, Germany
   \email{\{timotheus.kampik,a.luecke,joern.horstmann,\\mark.wheeler,david.eickhoff\}@sap.com} 
}
\authorrunning{Kampik et al.}
\titlerunning{SIGNAL}
\maketitle
\begin{abstract}
This paper provides an introduction to and discussion of SIGNAL, an industry-scale process data querying language and engine for large-scale cloud-based systems that is developed by SAP Signavio.
SIGNAL is optimized for fast read access to process data in event log format and utilizes an in-memory columnar store to this end. To facilitate usability, SIGNAL uses an SQL-like syntax with additional domain-specific querying features and in particular row-pattern matching-based temporal operators.
Also, the paper highlights research challenges related to process querying that are informed by the implementation and application of SIGNAL.

\keywords{Business Process Management \and Process Mining \and Query Languages}
\end{abstract}
%
%
\section{Introduction}
\label{sec:intro}
A key task within business process management is the inference of actionable insights from process data and knowledge.
To increase technology support for this task, process querying has emerged as an important sub-field of academic study and industry practice~\cite{DBLP:books/sp/22/P2022}.
As the consolidation of business process management and analysis technologies in the industry leads to a maturing technology ecosystem tackling large-scale challenges, there is an increased interest in the alignment of the academic body of knowledge on process querying with engineering-oriented real-world perspectives.
To this end, this paper provides an introduction to and discussion of SIGNAL, the SAP Signavio Analytics Query Language that is used for process analysis and in particular for event log-based process mining at SAP.
SIGNAL is a query language, as well as the name of the query engine that underlies the language.
Although the syntax of SIGNAL is SQL-like (and, indeed, a subset of SQL with several extensions for working with nested event log data), the underlying engine features a columnar database that is optimized for fast read operations in cloud-based multi-tenant environments.
In contrast to querying languages that elevate temporal/control flow-specific features that are characteristic for the process domain to operator level, SIGNAL uses regular expression based-querying for matching cases to this end, analogously to row pattern matching as introduced in the 2016 update of the SQL standard~\cite{10.1145/3299887.3299897} and somewhat similarly to some features of the \emph{Process Querying Language} (PQL)~\cite{DBLP:conf/bpm/PolyvyanyyCCRRF15,DBLP:journals/is/PolyvyanyyPH20}. This ensures a separation of concerns between traditional SQL features and process querying-specific extensions.
In this paper, we provide an introduction to SIGNAL, focusing on its motivation and design principles (Section~\ref{sec:purpose-principles}), its architecture (Section~\ref{sec:arch}), and the key query language features and how they relate to the underlying database (Section~\ref{sec:features}).
We also discuss SIGNAL in the context of other process query languages and under consideration of broader querying and database technology trends to then highlight long-term challenges that we see related to industry-scale process querying languages and engines (Section~\ref{sec:discussion}) before we conclude the paper (Section~\ref{sec:conclusion}).

\section{Process Querying}
\label{sec:process-querying}
%
The notion of a business process is a powerful abstraction for managing how work is conducted in (and across) organizations.
To facilitate the management of business processes, a range of technologies for process design, execution, and analysis exist.
A key to these technologies is the extraction, transformation, and analysis of process data and knowledge, often in the form of event logs~\cite{DBLP:books/sp/Aalst16} (data-driven) and process models~\cite{DBLP:books/sp/Weske19} (knowledge-based).
During the advent of business process management software, these tasks were handled by general-purpose database technology and typically by Structured Query Language (SQL)-based tools that were popular at the time, such as MySQL\footnote{For example, the APROMORE platform was (at least initially) implemented with an underlying MySQL database~\cite{LAROSA20117029}. However, let us note that some of the technology-centered assessments and claims in this paper are based on tacit engineering knowledge, in some cases of proprietary technologies, and hence cannot be traced back to academic papers or even reliably available open source code repositories.}.
However, over time, process querying technology has become increasingly sophisticated.
This development has been facilitated by two trends:
\begin{enumerate}
    \item Generally, SQL-based, relational databases were called into question regarding their ability to handle large amounts of data in large-scale and often cloud-based distributed systems. This ultimately led to the emergence of so-called \emph{NoSQL} database technologies that relax traditional constraints~\cite{CORBELLINI20171}. However, from a practical engineering perspective one may argue that the wide-spread adoption of some NoSQL technologies and associated challenges have also highlighted the fact that moving away from SQL is not a silver bullet. Consequently, engineers have started to carefully deliberate the trade-offs that are necessary when choosing between different database systems or when selecting features of new database systems that are to be implemented~\cite{ROYHUBARA2022101950}.
    \item Specifically, process management technologies have evolved, both in terms of their practical scale and their methodological sophistication. Regarding the former dimension, business process management software, and in particular data-intense process mining tools, have by now been consolidated into the product and application ecosystem of most of the large enterprise software vendors. Regarding the latter, the applications of process management software have been extended, with a focus change from modeling and execution to advanced mining and analysis approaches that often apply special-purpose algorithms, \emph{e.g.} for the discovery of process models from data or (temporal) conformance checking.
\end{enumerate}
The second trend is reflected in the business process management academic community by an analogous change from process modeling to mining, and to the development of process querying methods that increasingly move emphasis from the former to the latter.
However, the researchers' typical objectives are, for example, the application of new fundamental analysis methods~\cite{DBLP:books/sp/22/MurillasRA22}; arguably, real-world engineering concerns for querying processes at scale are rarely considered\footnote{A notable exception is a book chapter describing the Celonis process query language~\cite{DBLP:books/sp/22/0001ABSGK22}, which we cover in the discussion (Section~\ref{sec:discussion}) and whose description does not cover many of the engineering perspectives on process querying and general database technologies that we provide in this paper.}.
Some academic approaches to process data querying aim to enable the handling of ``big data'' (e.g.~\cite{DBLP:books/sp/22/BeheshtiBMGA22}). However, their architectures -- which are, for example, centered around technologies like SparkQL and Hadoop -- do not reflect the requirements for querying process data (and in particular event logs with tens of millions of rows per logs) of thousands of organizations simultaneously in cloud-based systems.
By describing and discussing SIGNAL, the process querying language and engine of SAP Signavio, this paper aims to shed more light on these practical considerations.

\section{Purpose and Principles}
\label{sec:purpose-principles}
\begin{figure}[ht]
\centering
\includegraphics[width=0.9\textwidth]{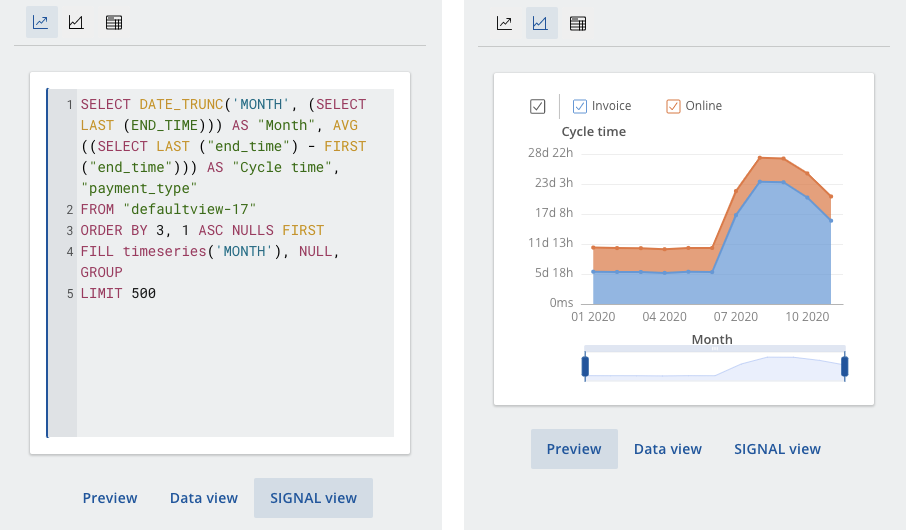}
\caption{A cycle time widget in preview mode and its underlying SIGNAL query.}
\label{fig:signal-chart}
\end{figure}

%
The main purpose of SIGNAL is to facilitate the analysis of event logs\footnote{Recall that an event log is a set of events and traditionally, each event has, at least, an event ID, a case ID, and a timestamp. Event logs are the classical data input for process mining algorithms and the most basic event log format (a set of (event ID, case ID, timestamp)-triplets) is sufficient for applying core algorithms for process discovery and conformance checking.}.
In particular, SIGNAL queries form the central abstractions of all event log analyses that are conducted through SAP Signavio's process mining software.
For example, business user-friendly process analysis ``widgets'' are, in fact, graphical user interfaces that trigger an underlying template-based instantiation of SIGNAL queries (see Figure~\ref{fig:signal-chart}) and SIGNAL queries can potentially be used by 3rd-party systems for the \emph{headless} usage of process mining capabilities.

Because modern enterprise software is typically delivered as-a-service, SIGNAL's design principles reflect engineering and product requirements for highly scalable, multi-tenant cloud-based software systems.
In particular, the following principles were given priority during query language and engine design.
\begin{description}
    \item[Cloud-based.]  The engine should be designed to run on hyperscaler environments for cloud-based systems.
    \item[Multi-tenant.] The underlying architecture needs to support a strict separation of tenants, \emph{i.e.}, environments of different customer organizations.
    \item[Serverless.] The query language and engine needs to be \emph{serverless} from the perspective of the user (SIGNAL users are data engineers, developers, and technical data analysts): users can focus on the central task of query-based inference, whereas technical details such as resource allocation, optimization, and execution are taken care of by the underlying system.
    \item[Mining-focused.] The language's primary purpose is the analysis of process data. Accordingly, the language should provide first-class abstractions for typical process mining tasks such as filtering for behavioral patterns, conformance checking, and process performance indicator (PPI) analysis; still, it must support mundane data analysis tasks as well.
    \item[Business user-friendly.] The query language must be relatively easy-to-use for business users with only basic programming skills. 
\end{description}
These key principles motivate SIGNAL's architectures and features, as presented in the following two sections.

\section{Architecture}
\label{sec:arch}
%
\begin{figure}[ht]
\centering
\includegraphics[width=0.65\textwidth]{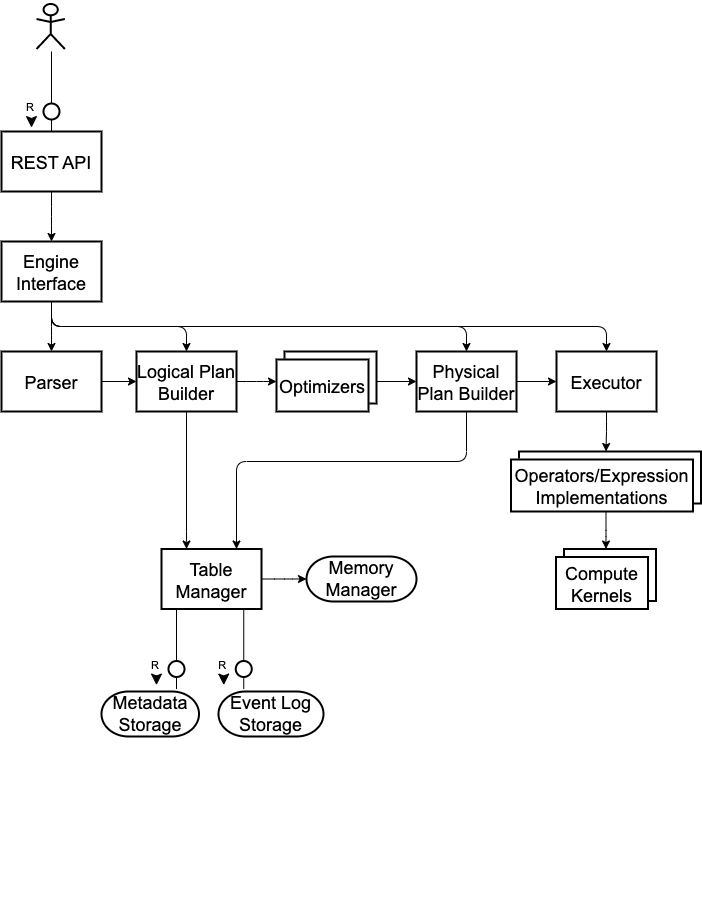}
\caption{SIGNAL engine architecture.}
\label{fig:signal-layers}
\end{figure}

SIGNAL's architecture is comprised of the following components (also see Figure~\ref{fig:signal-layers}).
\begin{description}
    \item[REST API and engine interface.] The SIGNAL query engine interface is exposed via a RESTful application programming interface (API) that allows users (humans or systems) to register SIGNAL queries and have the engine execute them and return the result. The main process mining tool that SIGNAL is integrated with uses SIGNAL via the REST API as the central language for producing process mining analyses. Users can either use SIGNAL directly (here, the tool abstracts from known contexts such as the current process scope) or use so-called \emph{widgets} in a graphical user interface that are then compiled to SIGNAL queries by the front-end system.
In addition to the REST API, a data ingestion API allows for the one-off upload of data, as well as for the regular ingestion of process data updates.
Supported formats are extensible event streaming (XES)~\cite{DBLP:journals/cim/AcamporaVSAGV17} and comma-/tab-separated value files (CSVs and TSV).
To facilitate the implementation of process mining and solutions, advanced data on-boarding tooling is available as part of the technology ecosystem that surrounds SIGNAL.
    
    \item[Parser.] Based on the SIGNAL language definition (described in Section~\ref{sec:features}), SIGNAL queries are parsed and the resulting abstract syntax tree is handed over to the logical plan builder.
    \item[Logical plan builder.] The logical plan builder takes the parsed query, executes a type check, and optimizes the query execution plan.
    On the logical level, a separation between case level and event level exists. On either level, the types \texttt{Boolean}, \texttt{Number}, \texttt{String}, \texttt{Timestamp}, and \texttt{Duration} are supported.
    \item[Optimizers.] Workloads are dynamically scheduled and optimized. Here, several of the optimization steps take advantage of the pre-sorted storage of the event logs (total order of events). This applies, for example, to distinct counts and window functions/partitions. To optimize for interactive performance, the \texttt{LIMIT} operator reduces the size of data to process as early as possible. This is especially useful for facilitating the analysis of process data through a browser-based user interface, where explorative queries are typically performed on a limited subset of the data by design.
    \item[Physical Plan Builder.] After optimization, the logical plan is converted into a physical execution plain. During the conversion snapshots of the required table columns are made. The physical plan is a tree or list of operators and executed in a separate thread to isolate effects of unexpected behaviors that may negatively impact performance.
    \item[Executor with operator and expression implementations, and compute kernels. \phantom{SIGNAL}] SIGNAL supports the typical operators and functions for SQL-based data analysis, plus additional operators for analyzing (temporal) control flow.
    The low-level compute kernels for relational algebra operations are optimized for high-throughput processing and work directly on the columnar data.
    \item[Table manager and memory manager.]  Memory and table management is enabled by open source frameworks for language-independent in-memory analytics using a columnar approach to data management that allows for the nesting of cases into event log tables.
    \item[Metadata storage and event log storage.] Compute and storage infrastructure is provided by different hyperscalers. Metadata is stored in a traditional relational database, whereas event logs are stored separately in the Apache Arrow columnar format\footnote{\url{https://arrow.apache.org/docs/format/Columnar.html}, accessed 30-01-2022.}, which allows fast read access because no complex joins are required.
\end{description}
Note that we keep the architecture description deliberately abstract, as implementation details, such as the exact technologies used for a specific layer, are prone to change and in some cases, different instances may run on somewhat different technology stacks.

A key aspect of SIGNAL is the columnar approach for data persistence.
As depicted in Table~\ref{fig:signal-nesting}, SIGNAL represents an event log as a table of cases, whereby each case has a nested table containing the events comprising the case. A case has a required \texttt{case\_id} attribute, each event has required \texttt{event\_name} and \texttt{end\_time} attributes, as well as an optional \texttt{start\_time} attribute\footnote{In our examples, we merely use \texttt{end\_time}, which reflects a typical real-world property of event logs: in most scenarios, an event has exactly one timestamp, representing the time at which the event has been registered in the system.}. Both cases and events can have several additional attributes (which are typically domain-specific). The events in each case are ordered by their \texttt{end\_time} timestamps.
This facilitates fast read operations.

\begin{table}
\centering
\caption{Columnar model for representing event logs.}
\label{fig:signal-nesting}
\begin{tabular}{ |c|c|c|@{}c@{}| }
\hline
\textbf{case\_ID} & \textbf{customer\_ID}  & \textbf{final\_status} & \textbf{events} \\
\hline
1001 & C2001 & \texttt{done} &
    \begin{tabular}{c|c|c} \textbf{event\_name} & \textbf{end\_time} & \textbf{status} \\\hline \texttt{Open ticket} & 1675086864052 & \texttt{none} \hspace{1.25pt}  \\ \hline \texttt{Assign ticket} & 1675160180724 & \texttt{open} \hspace{1.25pt} \\\hline \texttt{Close ticket} & 1675220315296 & \phantom{e} \texttt{done} \phantom{e} \hspace{1.25pt} \\\end{tabular} \\
\hline
1002 & C2002 & \texttt{blocked} &
 \begin{tabular}{c|c|c} \textbf{event\_name} & \textbf{end\_time} & \textbf{status} \\\hline \texttt{Open ticket} & 1675147138009 & \texttt{none} \\ \hline \texttt{Assign ticket} & 1675213914098 & \texttt{open} \\\hline \texttt{Close ticket} & 1675282027657 & \texttt{blocked} \\
 \hline \texttt{Open ticket} & 1675414104525 & \texttt{blocked} \\ 
 \end{tabular} \\
\hline
\end{tabular}
\end{table}

\section{The Query Language}
\label{sec:features}
%
The SIGNAL language is designed to query event logs represented in columnar format as described above (Figure~\ref{fig:signal-nesting}), using an SQL-based syntax, with a focus on read operations.
Here, the table that is queried from is typically fixed, \emph{i.e.}, a mere pointer by identifier to the corresponding ingested event log has to be provided; in many scenarios, this is handled by the integrating system, for example by a front-end application for process analysis.
In addition to standard SQL operators for data extraction and aggregation, SIGNAL features a row pattern matching approach for managing logical time (orders), \emph{i.e.}, for identifying cases with behaviors (events) matching the specified temporal constraints.

Before going into the details of temporal pattern matching with SIGNAL, let us provide some remarks regarding the assumptions we make about information that is available on case and event level, and in particular about the availability of timestamps.
A minimal case definition entails a case identifier, a sorted list of event names, event start timestamps (empty if missing), and event end timestamps.
This means the start timestamps of events in event logs yield a partial preoder on the events, whereas the end timestamps are guaranteed to yield a total preorder -- however, because on event log-level, a sequence is enforced and ties are broken in case of non-unique timestamps, we can assume a partial order and a total order, respectively.
In practice event start timestamps are typically missing, as the execution times of activities are often not logged; in the enterprise systems that are the ultimate sources of event logs, an event is typically created only when an activity is concluded.
For example, when a purchase order is created, there is typically no start time; merely the submission of the form triggers an event.
Indeed, one could argue that even if opening the form logs a start timestamp, this timestamp is potentially misleading:
after all, the user may have spent substantial time preparing the purchase order before opening the form.
In case users want to rely on classical SQL only (without row pattern matching, but with some additional basic operators for handling temporal relationships between events), they can flatten the case tables that are nested into the event log table.

\begin{figure}[ht]
\centering
\includegraphics[width=1\textwidth]{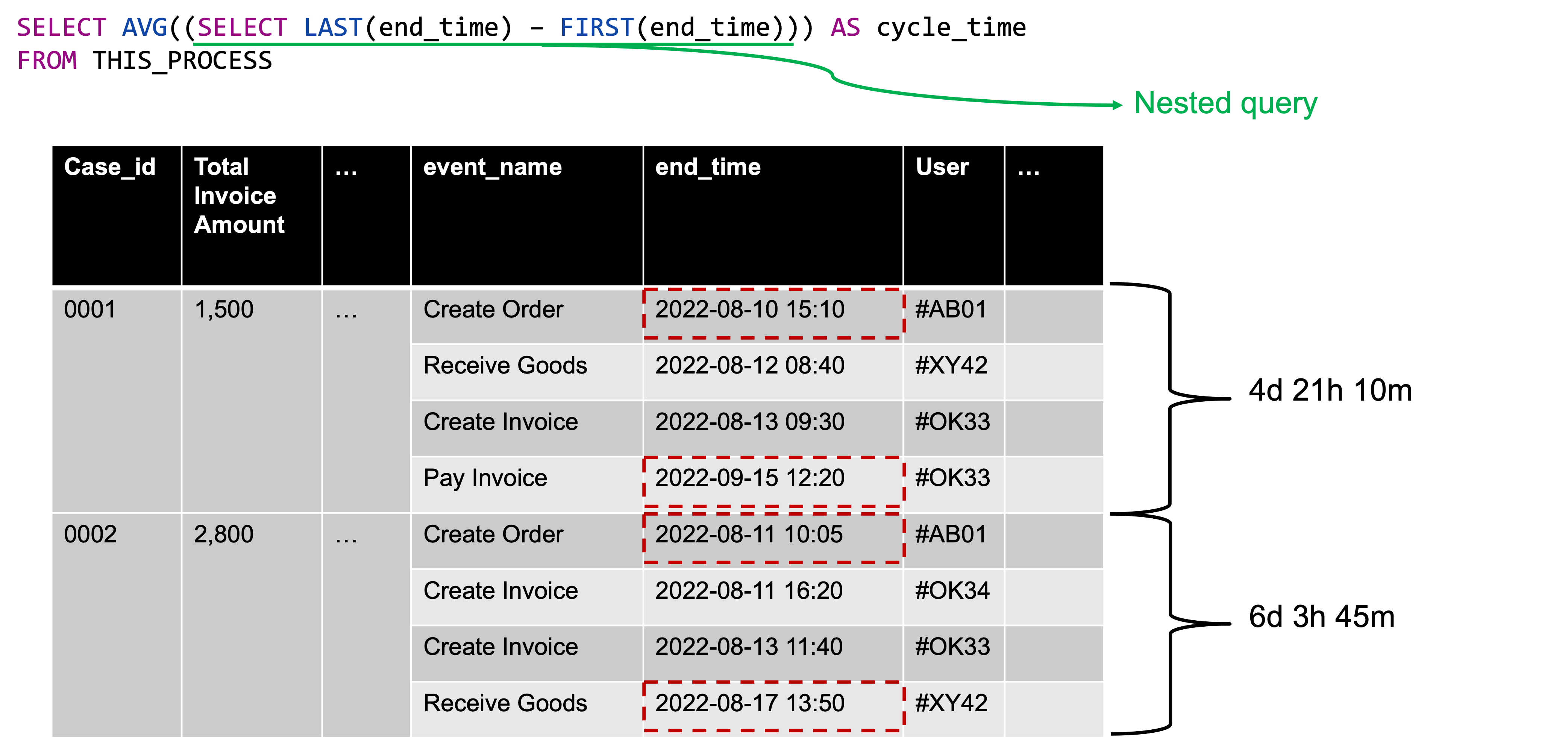}
\caption{Querying nested tables with SIGNAL: average cycle time. The selected cycle times (from which the average is then computed) are displayed on the right hand-side.}
\label{fig:nested-query}
\end{figure}

Figure~\ref{fig:nested-query} shows a nested SIGNAL query that obtains the average cycle time of a process from an event log persisted in the SIGNAL database and visualizes how the result is obtained.
Note that row pattern matching is not used; instead, the query relies on the \texttt{FIRST} and \texttt{LAST} operators.
The event-level sub-query \texttt{(SELECT LAST(end\_time) - FIRST(end\_time)} can be thought of as having an implicit \texttt{FROM events} clause. This means that logically, expressions inside the aggregation operate on individual rows of the nested \texttt{EVENTS} table.
\texttt{FIRST} and \texttt{LAST} are aggregation functions (just like \texttt{MIN} or \texttt{MAX}) that reduce several inputs to a single value.
The result of event-level sub-queries must be a single value, which we can ensure by requiring an aggregation function inside the sub-query.
The event-level sub-query is important from a type-checking perspective. Outside of it, an expression involving \texttt{end\_time} would refer to multiple values, which is not allowed.

The behavior of the row pattern matching approach can be demonstrated in a semi-formal way using a simple graph-based model of a case.
We define a case $C = (E, R)$ as a finite set of events $E$ and $R \subseteq E \times E$, where we denote $(a, b) \in R$ by $a \succeq b$ and $(a, b) \in R, (b, a) \not \in R$ by $a \succ b$,
and stipulate that $R$ is a total order\footnote{Recall that because $R$ is a total order, $\forall a, b, c \in R$ it holds that $a \succeq a$ (reflexivity), if $a \succeq b$ and $b \succeq c$ then $a \succeq c$ (transitivity), and if $a \succeq b$ and $b \succeq a$ then $a = b$, as well as $a \succeq b$ or $b \succeq a$; note that we have a total order and not a total preorder, even in case of non-unique timestamps because the order is ``burned in'' based on the time-ordered sequence of events in the event log, with ties broken for non-unique timestamps.} on $E$ that can be inferred from the end timestamps of the case events.
To check whether a case satisfies temporal constraints, we introduce a finite set of behavior constraints $B$, as well as a behavioral matching function (partial function) $f: B \rightarrow 2^E$ that maps behaviors to sets of events.

Let us provide a brief SIGNAL language tutorial that first gives some general intuitions before we define the row pattern matching approach more precisely.
For example, in a customer support request handling process, we may want to define cases in which tickets have been re-opened after they have been closed. 
With SIGNAL, we can query this behavior (with some assumptions made about the provided event names) as follows.
\begin{minted}[escapeinside=||,mathescape=true,tabsize=4,obeytabs, fontsize=\footnotesize]{sql}
    SELECT case_id
    FROM THIS_PROCESS
    WHERE event_name MATCHES ('Close ticket' |$\sim >$| 'Open Ticket')
\end{minted}
Note that \texttt{THIS\_PROCESS} is a context-dependent pointer to the process that is currently in scope, \emph{e.g.}, in the context of a visual data analysis as executed through a graphical user interface by a business user. \newpage

The \texttt{MATCHES} clause works as follows:
\begin{itemize}
    \item The scope of a match is a case (\emph{i.e.}, a set of events) and not a single event, as specified by our behavioral matching function $f$.
    \item Our \emph{behaviors} are implicitly defined by \texttt{event\_name} (a column name) which is then assigned the values \texttt{'Close ticket'} and \texttt{'Open ticket'}. Behaviors can be defined explicitly, by selecting events that satisfy specified constraints, as can be seen in the examples further below. Given the implicit definition, all events with the names \texttt{'Close ticket'} and \texttt{'Open ticket'}, respectively, are matched.
    \item The temporal pattern matching operator \texttt{$\sim >$} (\emph{directly or indirectly followed by})\footnote{\texttt{$\sim >$} is more precisely explained below.} then matches cases in which behaviors with the specified temporal relationship occur.
    In this example, all cases in which an event for which \texttt{event\_name = 'Close ticket'} holds is indirectly followed by an event for which \texttt{event\_name = 'Open ticket'} holds.
\end{itemize}

Also, we want to identify cases in which tickets are closed although they are marked as ``blocked'' (instead of the expected status ``done'' that typically should be set when a ticket is closed).
With SIGNAL, we can query this behavior (with some \emph{additional} assumptions made about the provided attributes) as follows.
\begin{minted}[escapeinside=||,mathescape=true,tabsize=4,obeytabs, fontsize=\footnotesize]{sql}
    SELECT case_id
    FROM THIS_PROCESS
    WHERE (event_name = 'Close ticket' AND "status" = 'blocked')
\end{minted}

Finally, let us identify the cases in which blocked tickets are first directly closed and then (eventually) re-opened.
Here, we make use of the notion of \emph{behaviors}. Note that arbitrary expressions (that return Boolean values) can be used to define a behaviors.
\begin{minted}[escapeinside=||,mathescape=true,tabsize=4,obeytabs, fontsize=\footnotesize]{sql}
    SELECT case_id
    FROM THIS_PROCESS
    WHERE BEHAVIOUR
    (event_name = 'Close ticket' AND "status" = 'blocked'))
    as closed_while_blocked
    MATCHES(closed_while_blocked |$\sim >$| 'Open ticket')
\end{minted}
In our framework, we can then model the latter query as follows:
\begin{itemize}
    \item The set of behaviors is $B = \{x, y\}$, where $x$ represents \\ \texttt{closed\_while\_blocked} and $y$ represents events with the activity name ``Open ticket''.
    \item The events in our case are $E = \{e_0, ..., e_n\}$ and we assume that they occur in a total order, ordered by index, \emph{i.e.}, we have a graph $C = (E, R)$, where $R = \{(e_i, e_j) | 0 \leq i \leq j \leq n \}$ (meaning $(E, R)$ is equivalent to the sequence $\langle e_0, ..., e_n \rangle$). 
    \item We have exactly one event $e \in E$ such that $e$ is a ticket that has been closed while in status ``blocked''; we have exactly two events $e', e'' \in E$ that have the activity name ``Open ticket'': $f = \{(x, \{e\}), (y, \{e', e''\})\}$. Let us assume that $e' \succ e$ and $e \succ e''$ hold.
    \item The matching operator \texttt{$\sim >$} (follows directly or indirectly) is then translated into the statement $\exists c \in f(x), d \in f(y): c \succ d$. This is the case: $e \in f(x), e'' \in f(y), e \succ e''$. Hence, our query matches the case.
\end{itemize}
Note that for any operator, matching is existential (and not universal), \emph{i.e.}, one match of behaviors is sufficient, as exemplified above.

Let us claim, based on well-established work, that using a single binary relation ($R$) for managing finite traces is sufficient for expressing linear temporal logic for finite traces~\cite{DBLP:conf/ijcai/GiacomoV13}.
Below, we assume a case $C = (E, R)$ with $a, b \in E$ and a set of behaviors $B$ with behaviors $a', b' \in B$ such that $a'$ and $b'$ match unique events, \emph{i.e.}, we have a behavioral matching function $f$ for which $f(a') = \{a\}$ and $ f(b') = \{b\}$ hold.

The following row pattern matching operators are available.
\begin{description}
    \item[Directly followed by] (\texttt{$->$}). Two events that match the behavior directly follow each other. $a'$ \texttt{$->$} $b'$ matches our case iff $a \succ b$ and $\nexists c \in E$ such that $c \neq a$,  $c \neq b$,  $a \succ c$, and $c \succ b$ hold.
    \item[Indirectly or directly followed by] (\texttt{$\sim >$}).
    Two events that match the behavior follow each other, directly or indirectly. As sketched above, $a'$ \texttt{$\sim >$} $b'$ matches our case iff $a \succ b$ holds.
    \item[Starts with] (\texttt{\^}). An event that matches the behavior is the first event in the case. \texttt{\^} $a'$ matches our case iff $\forall c \in E \setminus \{a\}$ it holds that $a \succ c$.
    \item[Ends with] (\texttt{\$}). An event that matches the behavior is the last event in the case. $a'$ \texttt{\$} matches our case iff $\forall c \in E \setminus \{a\}$ it holds that $c \succ a$.
    \item[Contains any] (\texttt{ANY}). Specifies that any event needs to occur. $a'$ \texttt{ANY} $b'$ matches our case iff $\exists c \in E$ such that for some $c' \in B: c \in f(c')$ it holds that $a'$ \texttt{$->$} $c'$ and $c'$ \texttt{$->$} $b'$.
    \item[Does not contain] (\texttt{NOT}). Negates a behavioral match. For example, $a'$ \texttt{NOT} $b'$ matches our case iff $\exists c \in E$ such that  $a \succ c$ and $\nexists d \in E$ such that $d \neq a$,  $d \neq c$,  $a \succ d$ and $d \succ c$, and $c \not \in f(b')$.
    \item[Or] (\texttt{$|$}). Either of two behaviors occur. For example, \texttt{\^} $(a'| b')$ matches our case iff \texttt{\^} $a'$ or \texttt{\^} $b'$ matches our case.
    \item[Repetition] (\texttt{*}). Specifies that a behavior occurs zero or more times. For example, $a'$ \texttt{ANY*} $b'$ matches our case iff $a'$ \texttt{$\sim >$} $b'$.
\end{description}
Let us highlight that universal quantification can be achieved by negation.
For example, if we want to express that behavior $a'$ should always be directly followed by behavior $b'$ we can stipulate this with the expression \texttt{ (\^~$($NOT $a'$ | ($a'$ $b'$)$)^*$ \$)}, which is then matched iff the whole sequence imposed by the total order on the events consists of anything that is not $a'$ , or $a'$ followed by $b'$., \emph{i.e.}, iff $a'$ occurs it is always directly followed by $b'$.
Table~\ref{tab:examples} illustrates how the pattern matching operators work, assuming again a total order (sequence) of events and the behaviors and behavioral matching function from the example above.
\begin{table}
\renewcommand*{\arraystretch}{1.5}
\centering
\begin{tabular}{ | c | c | }
 \hline
 \thead{Pattern Syntax} & \thead{Matching Sequence} \\
  \hline
 $a'$ \texttt{$->$} $b'$ & $\langle ..., a, b, ... \rangle$  \\ 
  $a'$ \texttt{$\sim >$} $b'$ & $\langle ..., a, ...,  b, ... \rangle$ \\ 
 \texttt{\^} $a'$ & $\langle a, ... \rangle$ \\ 
  $a'$ \texttt{\$} & $\langle ..., a \rangle$ \\ 
 $a'$ \texttt{ANY} $b'$ & \makecell{$\langle ..., a, c, b, ... \rangle$ \\ (given some event $c \in E$)} \\ 
 $a'$ \texttt{NOT} $b'$ & \makecell{$\langle ..., a, c ... \rangle$ \\ (given some event $c \in E: c \not \in f(b')$)} \\
 \texttt{\^} $(a'| b')$ & \makecell{$\langle c, ... \rangle$ \\ (given $c \in \{a, b\}$)} \\ 
 $a'$ \texttt{ANY*} $b'$ & $\langle ..., a, ...,  b, ... \rangle$ \\ 
 \hline
\end{tabular}
\caption{Pattern matching operators and sequences (total orders) matched by them.}
 \label{tab:examples}
\end{table}

\begin{figure}[ht]
\centering
\includegraphics[width=0.7\textwidth]{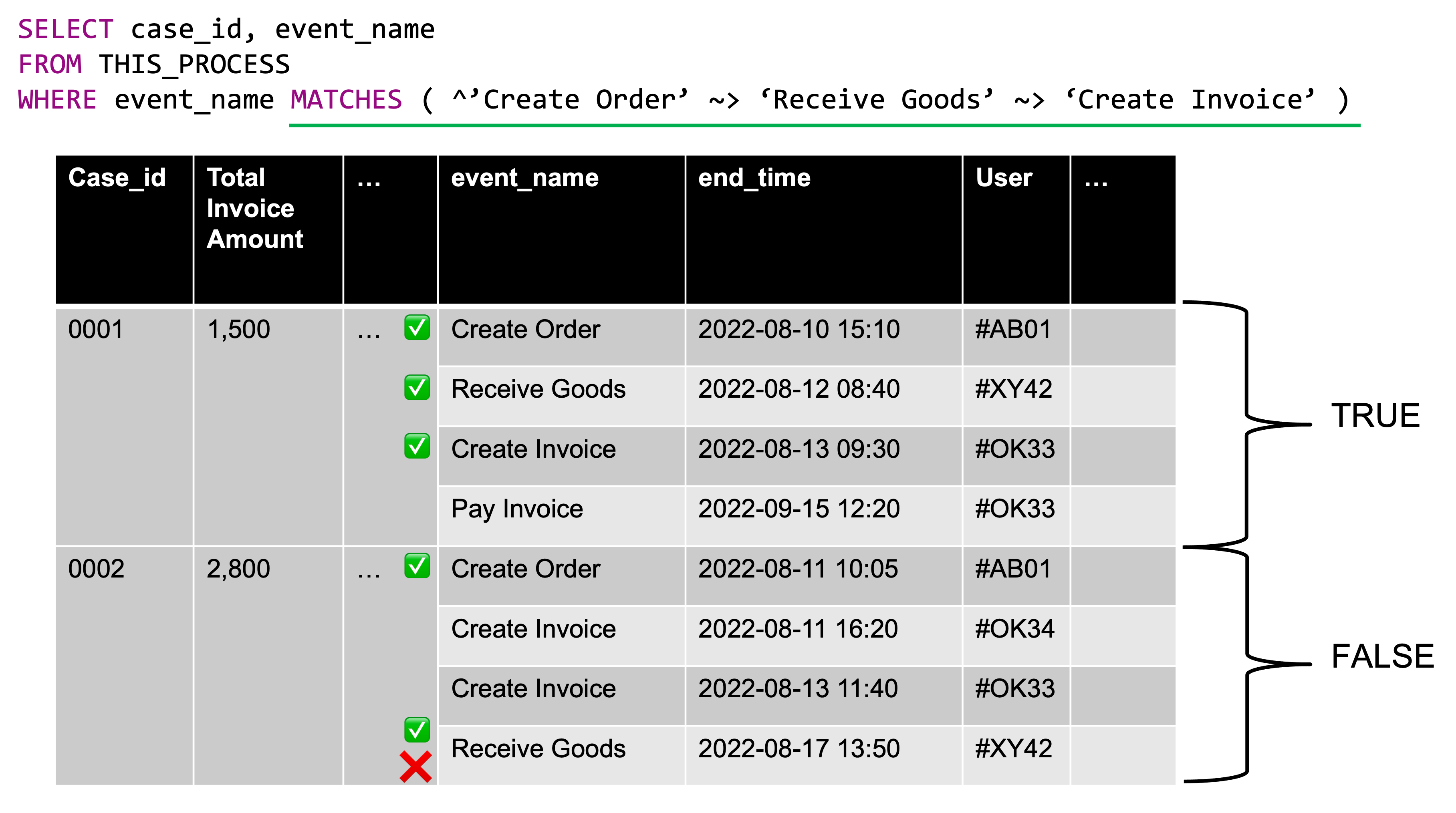}
\caption{Applying control flow constraints with SIGNAL: the bracket labels on the right hand-side indicate which cases match the constraints. }
\label{fig:signal-constraints}
\end{figure}

Figure~\ref{fig:signal-constraints} shows a query in which control flow constraints are specified using regular expressions for row pattern matching.
In particular, the activity ``Create Order'' must occur first in the ordered (by timestamp) sequence of events for a particular case, followed (directly or indirectly) by ``Create Order'' , which in turn must be followed (directly or indirectly) by ``Create Invoice''.

%

\section{Discussion}
\label{sec:discussion}
%
To the best of our knowledge, SIGNAL is the second commercial, industry-scale process querying language that is presented in the academic literature; the only language that has been presented before is the Celonis Process Querying Language (CPQL)~\cite{DBLP:books/sp/22/0001ABSGK22}.
There are several similarities between SIGNAL and CPQL.
Obviously, SIGNAL and CPQL cover similar use cases, focusing on large-scale process mining support.
Also, both languages make use of in-memory databases and focus on fast read operations and custom operators for managing logical time.
However, in contrast to CPQL, SIGNAL makes use of a columnar approach to data storage to facilitate fast read access (CPQL is reported to rely on a snowflake schema) and the SIGNAL syntax has a clear separation of concerns between traditional SQL operators and custom temporal operators for event log analysis, (the latter of which is handled with row pattern matching).
Also, while CPQL is reported to be ``inspired by SQL'', SIGNAL is arguably closer to the SQL model, extending it with the concept of a nested events table. On this nested table SIGNAL can operate with the same SQL operations and aggregations as on case-level. In contrast, CPQL requires special syntax or functions to pull, for example, aggregated values from an activities table to case-level.
A particularly noteworthy difference between SIGNAL and CPQL is that reportedly, CPQL relies on Petri Net-based representations for process conformance checking; in contrast, SIGNAL takes the principle-based stance that its row pattern matching-based temporal inference are the only first-class abstraction for temporal reasoning, assuming that this leads to a simpler, more maintainable, and ultimately more performant and user-friendly language and query engine.
To the best of our knowledge SIGNAL is, at the time of writing, unique also when compared to academic process querying languages and engines in its use of a columnar database and SQL-based syntax with row pattern-matching extension for managing logical time.
These key design decisions can potentially have interesting implications for academic research: in the SAP Signavio process mining ecosystem, all event log analysis tools and means, like graphical user interface widgets, are in the end compiled to SIGNAL queries.
In academic as well as in broader open technology ecosystems, a similar approach, focusing on a potentially standardized lingua franca could help to further channel efforts around process mining and facilitate long-term innovation, in particular when considering how other database technologies such as Apache Flink~\cite{DBLP:journals/debu/CarboneKEMHT15} have emerged (at least partially) from academia.

Based on experiences with implementing and building solutions for SIGNAL, we identify the following trends and challenges for modern, industry-scale process query languages and engines.
\begin{description}
    \item[Support during data transformation.]
    A key challenge in process mining is the extraction of event logs from enterprise software databases~\cite{DBLP:conf/bpmds/KampikW22}.
    Typically, the ETL process is complex and iterative and requires, already prior to a process mining-based analysis, an advanced understanding of a business process, the involved IT systems, and the data that they generate.
    Process querying languages can potentially play a role already during the ETL process, for example when fusing different data sources (as described in the previous section).
    As a prerequisite for further exploring this direction, future studies may, for example, investigate the challenges of data fusion in process mining application scenarios.
    \item[Event log-less process data analysis.]
    Going one step further, process data can ideally be queried directly from the databases of one or several IT systems.
    This is, in many scenarios, not a realistic approach, for example because the underlying databases are not sufficiently performant or because the domain knowledge that is required to move from the process data as represented in the database to a representation that can be analyzed with reasonable ease renders an intermediate transformation inevitable.
    However, in many scenarios, the data transformation challenge requires in-depth source system database expertise in any case and performance bottlenecks can potentially be offset by persisting process mining-friendly \emph{views} on enterprise system data.
    Hence, process data querying can be expected to move closer to the source system in the future.
    \item[Multiple views, depending on different notions of case ID.]
    From a practical perspective, a traditional, flat event log provides merely one particular view on a business process and depends on the scoping of a process according to a particular case ID notion.
    For example, an order handling process may use the order ID, the order item ID, or the shipment ID as case ID, and the resulting event logs allow for vastly different perspectives on how orders are handled, focusing on items with different, but not necessarily hierarchically related granularity levels (for example, an order may be split across several shipments, but a shipment may also contain items of several orders).
    Academically, this challenge is addressed by the introduction of object-centric process mining~\cite{DBLP:conf/sefm/Aalst19}, adding object-oriented abstractions to the event log (which is then no longer a simple table).
    However, one may argue that object-centric process mining merely moves more structure from the relational databases that event logs are typically extracted from into the event log itself.
    Possibly, an alternative approach could be to provide dynamic process scoping support during the ETL process and then implement first-class abstractions in process querying languages that can explore several scopes (case ID notions) for one set of process data. For example, in the context of order-to-delivery process, one can explore the scope of an order, a delivery, or even the journey of a particular purchase item.
    \item[Integration of mining and modeling.] SIGNAL focuses entirely on querying event logs (process data), although in its product context, process models (\emph{i.e.}, representations of process models) are typically available.
    Potentially, SIGNAL can also be used to query process models.
    However, process model collections are typically relatively small (compared to event logs or collections thereof) and the temporal ordering of activities in models (e.g.: ``return all process models where creating a purchase order is followed by an approval step'') is rarely important when querying; \emph{i.e.}, although we consider that process model querying can have a substantial business impact, for example in the context of variant management and best practice recommendation, we assume that the precise querying of ``temporal'' control flow is of lesser importance here.
    \item[First-class abstractions for organization and experience mining.]
    Ultimately, process management is concerned with the management of organizations.
    However, key organizational perspectives, in particular roles and resources, are not treated as first-class abstractions in typical process querying languages.
    Still, a substantial body of research exists on \emph{organization mining}~\cite{DBLP:journals/dss/SongA08}\footnote{Originally referred to as \emph{organizational mining} but we argue that the mining is not \emph{organizational}; rather, the object that is mined is an \emph{organization}.}, the extraction and analysis of organizational information from event logs.
    More recently, a first approach to \emph{agent system mining}~\cite{DBLP:journals/access/TourPK21} has emerged, focusing on the organizational resource that executes particular activities as a first-class abstraction. In particular the latter approach raises the question to what extent process querying languages can handle organizational perspectives that are specific to particular actors, \emph{i.e.}, to the experiences (or: \emph{beliefs}) of human or artificial agents acting therein.
    For example, when process data is analyzed from several IT systems with inconsistent representations of the flow or state of work, then the inconsistency can be observed through queries and can allow for conclusions that agents that work with either the one or the other IT system hold false beliefs that may lead to incorrect decision-making; from a foundational perspective, such analyses can be aided by tools from non-classical logics, e.g., using paraconsistent or non-monotonic logic. 
    \item[Explainable inference.] Process querying can be considered a method of symbolic inference (from either data or knowledge). A recent trend in symbolic reasoning is the generation of (often concise, \emph{i.e.}, in some sense minimal) \emph{explanations}: queries or results thereof that provide a compelling summary of why a particular inference has been drawn, e.g., using notions of necessity, sufficiency, or counterfactuality, and can ideally be rendered into a human-interpretable form. Tapping into the wealth of reasoning explainability (see~\cite{DBLP:journals/corr/abs-2009-00418} for a survey) can potentially facilitate the application of process querying to key application areas of business process intelligence software such as root cause analysis and process change recommendation generation.
    \item[Cross event-log, cross-organization, and cross-process mining.]
    Similarly to the data fusion and process scoping challenges, process querying languages should ideally support joint inference from several event logs at once. Traditionally, the assumption is that the instances in the event logs that are to be jointly analyzed are connected. 
    However, jointly analyzing several implementations of the same process across organizations is a use case that is promising as well, in particular because it can drastically reduce analysis efforts for individual organizations, while at the same time facilitating correct and actionable analysis results.
\end{description}

\section{Conclusion}
\label{sec:conclusion}
%
In this paper, we have presented an industry-scale process querying language and engine.
We hope that we have shed more light on industry-scale approaches to process querying for event log analysis that can inform future research on the topic.
In particular, we think that a convergence of academic and industry-oriented approaches that consider practical constraints and challenges can fuel innovation on process querying and ideally even generate insights that are more broadly applicable to general database technologies.

\subsubsection*{Acknowledgments}
\label{sec:ack}
We thank the numerous engineers and analysts working on and with SIGNAL. We would also like to thank Artem Polyvyanyy, Adrian Rebmann, Han van der Aa, and Matthias Weidlich for in-depth discussions and feedback. 
%
%
%
\bibliographystyle{splncs04}
\bibliography{references}
\end{document}